\documentclass[%
reprint,
 amsmath,amssymb,
 aps,
prb,
]{revtex4-1}

\usepackage{graphicx}% Include figure files
\usepackage{dcolumn}% Align table columns on decimal point
\usepackage{bm}% bold math
\usepackage{xcolor}
\begin{document}

\preprint{APS/123-QED}

\title{
First-principles study of the charge ordered phase in $\kappa$-D$_3$(Cat-EDT-TTF/ST)$_2$:
Stability of $\pi$-electron deuterium coupled ordering in hydrogen-bonded molecular conductors
}
%}% Force line breaks with \\
%\thanks{A footnote to the article title}%

\author{Takao Tsumuraya}\thanks{E-mail: tsumu@kumamoto-u.ac.jp}
\affiliation{Priority Organization for Innovation and Excellence, Kumamoto University, 2-39-1 Kurokami, Kumamoto, 860-8555, Japan}
\author{Hitoshi Seo}
\affiliation{Condensed Matter Theory Laboratory, RIKEN, 2-1 Hirosawa, Wako, Saitama, 351-0198, Japan}
\affiliation{Center of Emergent Matter Science~(CEMS), RIKEN, 2-1 Hirosawa, Wako, Saitama, 351-0198, Japan}
\author{Tsuyoshi Miyazaki} 
\affiliation{International Center for Materials Nanoarchitectonics~(WPI-MANA), National Institute for Materials Science, 1-1 Namiki, Tsukuba, Ibaraki 305-0044, Japan}
%\collaboration{CLEO Collaboration}%\noaffiliation

\date{\today}% It is always \today, today,
             %  but any date may be explicitly specified

\begin{abstract}
~We study the electronic and structural properties of the low-temperature ordered phase of hydrogen-bonded molecular conductors, $\kappa$-D$_3$(Cat-EDT-TTF)$_2$ and its selenium-substituted analog $\kappa$-D$_3$(Cat-EDT-ST)$_2$, by means of first-principles density functional theory~(DFT) calculations.
In these compounds, the charge ordering in the $\pi$-electron system is coupled with the ordering of the displacements in the deuteriums forming the hydrogen-bond, equally shared by two oxygens in the high-temperature phase. 
While the structural optimization within the standard DFT method based on the generalized gradient approximation fails to reproduce the structural stability of the charge-ordered phase, we show that a hybrid functional of Heyd, Scuseria, and Ernzerhof can reproduce structural characters of the CO phase, owing to the more localized nature of the wave functions.
Furthermore, using the ability of the hybrid functional to predict the electronic and structural properties, we find a stable noncentrosymmetric charge ordered phase with another pattern of deuterium ordering. 
%\begin{description}
%\item[Usage]
%Secondary publications and information retrieval purposes.
%\item[PACS numbers]
%\pacs{31.15.E, 31.70.Ks, 31.10.+z}
%31.70.Ks Molecular solids
%31.10.+z Theory of electronic structure, electronic transitions, and chemical binding}+ command.
%\item[Structure]
%You may use the \texttt{description} environment to structure your abstract;
%use the optional argument of the \verb+\item+ command to give the category of each item. 
%\end{description}
\end{abstract}

\pacs{Valid PACS appear here}% PACS, the Physics and Astronomy
                             % Classification Scheme.
%\keywords{Suggested keywords}%Use showkeys class option if keyword
                              %display desired
\maketitle

%\tableofcontents

\section{\label{sec:level1}Introduction}
Accurate calculation of structural and electronic properties for crystalline solids hosting strongly correlated electron systems is a long-standing problem in condensed matter physics.
The standard first-principles calculations based on the density functional theory (DFT) which have shown great success in many materials
often fail to reproduce the insulating states such as Mott insulators\cite{Terakura_MnO_1984, Miyazaki99_dmit, k_ET_KNakamura_JPSJ09} and charge ordered (CO) insulators.~\cite{Yamauchi_PRB09, Yoshimi2012, Jacko_TMTTF_2013, Yamakawa_alpha_ET2I3, Katoh_NaV2O5, Anisimov_Fe3O4, Guo_Magnetite2004, Botana_La4Ni3O8}
One of the prototypical systems is molecular conductors in which correlated electrons are formed owing to comparable energy scales for the kinetic energy and the Coulomb interactions among electrons.\cite{SeoHott_ChmRev04}
A fundamental problem is that the CO phases in these molecular systems, often seen experimentally and extensively studied theoretically, are not structurally stable within the standard DFT approach. 
Namely, even when we perform structural optimization starting from the experimentally observed low-temperature structure with charge ordering, the optimized structure becomes that of the high-temperature phase where the charge disproportionation is absent. 

\begin{figure}[tb]
\begin{center}
\includegraphics[width=1.0\linewidth]{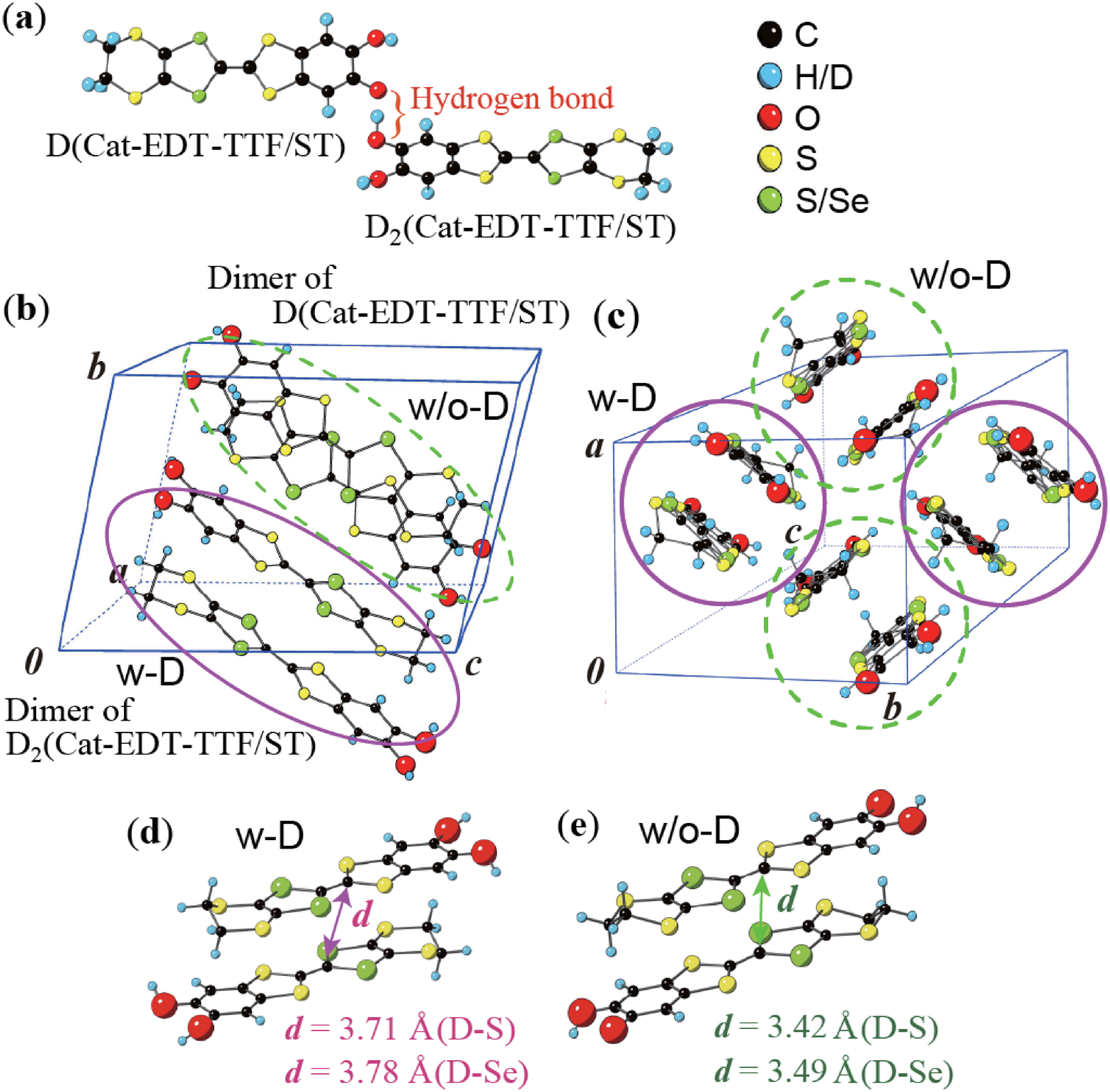}
\end{center}
\caption{(Color online) (a) Molecular structure of D$_3$(Cat-EDT-TTF/ST)$_2$. (b) Crystal structure of the low-temperature phase of $\kappa$-D$_3$(Cat-EDT-TTF/ST)$_2$ (space group: $P\bar{1}$)~\cite{A_Ueda_DCat_Eur}, and (c) its view along the $c$ axis showing the $ab$ plane. The solid (dashed) ellipses in (b) and circles in (c) indicate dimers of the molecular units with D (without D).
Their dimer structures: (d) D$_4$(Cat-EDT-TTF/ST)$_2$, and (e) D$_2$(Cat-EDT-TTF/ST)$_2$. Experimental values of intermolecular atomic distance between C atoms are shown for both D-S and D-Se.}
\setlength\abovecaptionskip{0pt}
\label{Crystal}
\end{figure}

Here we study a class of molecular crystals based on catechol with ethylenedithiote-tetrathiafulvalene, Cat-EDT-TTF, and its selenium-substituted analog, Cat-EDT-ST.~\cite{Kamo12_H3_Cat}  
These systems are characteristic in the sense that their electronic and structural properties are strongly linked, through the hydrogen bonding in their constituent molecular units.
The units form the so-called $\kappa$-type two-dimensional arrangement; the solids are called $\kappa$-H$_3$(Cat-EDT-TTF)$_2$~(hereafter abbreviated as H-S) and $\kappa$-H$_3$(Cat-EDT-ST)$_2$ (H-Se). 
Their deuterated samples, $\kappa$-D$_3$(Cat-EDT-TTF)$_2$ (D-S) and $\kappa$-D$_3$(Cat-EDT-ST)$_2$ (D-Se), are also prepared and experimentally investigated. 
Interestingly enough, large differences between the hydrogen and deuterium samples are seen. 
At high temperatures, they all show Mott insulating behavior with localized $S$ = 1/2 spins.~\cite{Isono_H3Cat_2013, Isono_QSL_H3Cat13} 
Structurally, every two H/D(Cat-EDT-TTF/ST) units share a H/D atom forming the hydrogen-bonding between two nearest oxygens, bridging the molecules~with a relatively short O$\cdots$O distance~[Fig.~\ref{Crystal}(a)].
These are consistent with the electronic structure calculations~\cite{Isono_H3Cat_2013, Tsumu_H3Cat} where the $\pi$-electrons form a half-filled band owing to the dimerization in the $\kappa$-type arrangement of H/D(Cat-EDT-TTF/ST) units which are all equivalent. 
The strong electron-electron repulsion is the probable reason the system is Mott insulating (the so-called dimer-Mott insulator\cite{SeoHott_ChmRev04, KinoFukuyama_dimer96, Miyazaki99_dmit, k_ET_KNakamura_JPSJ09}). 
Then the $S$ = 1/2 spins are localized on every dimer, namely the antibonding pair of the highest occupied molecular orbitals (HOMO) of H/D(Cat-EDT-TTF/ST) molecules.

The difference between the H and D samples appears at low temperatures. 
The H samples stay paramagnetic down to the lowest temperatures, showing the possibility of a quantum spin liquid state.~\cite{Isono_QSL_H3Cat13}  
The coupling between the quantum fluctuation of protons and $\pi$-electrons is discussed based on a dielectric constant measurement showing a quantum paraelectric behavior.~\cite{Shimozawa_H3Cat2017}
In fact, our previous first-principles DFT calculations show that the optimized distance between the two O atoms is relatively short, and the calculated potential energy surface for the shared H atom is very shallow near the minimum points.~\cite{Tsumu_H3Cat}  

In stark contrast, in the D samples, a first-order phase transition occurs that is associated with a large structural change at a transition temperature~($T_c$) of 182 and 185~K in D-S and D-Se, respectively.~\cite{A_Ueda_DCat, A_Ueda_DCat_Eur, Ueda_JPSJ2019} 
The magnetic susceptibility decreases sharply below $T_c$ resulting in a nonmagnetic ground state.~\cite{Shimozawa_H3Cat2017, SYamashita_DCat2017}
The heat capacity of D-S shows smaller values than in H-S at low temperatures, consistent with the absence of spin contribution.\cite{SYamashita_DCat2017}
Below $T_c$, the D atom forming the hydrogen bond localizes near one of the two O atoms. 
The low-temperature structure (space group: $P\bar{1}$) suggests the existence of charge disproportionation between two types of dimers that consist of two distinct molecules, D$_2$(Cat-EDT-TTF/ST) and D(Cat-EDT-TTF/ST), abbreviated here as w-D and w/o-D units, respectively~[Figs.~\ref{Crystal}(b) and (c)].~\cite{A_Ueda_DCat, A_Ueda_DCat_Eur} 
It is considered that the charge ordering in the $\pi$-electron system is coupled with the D ordering.~\cite{Naka_PRB_H3Cat18} 

In our previous study,~\cite{Tsumu_H3Cat} we performed structure optimization for H-S and found a H-coupled CO phase, which has the same symmetry as the low-temperature structure of D-S and D-Se. 
However, the results show a metallic state with a large Fermi surface
in contrast to the insulating behavior in experiments, and the magnitude of charge imbalance, and the structural distortions from the high-temperature phase are much smaller than those observed in the D samples. 
These calculations were done with a conventional exchange-correlation functional of the generalized gradient approximation (GGA) proposed by Perdew, Burke, and Ernzerhof (PBE),~\cite{GGA_PBE} which often underestimates the localization of electrons and then the structural stability of CO phases.
Since the band structure, distortion of molecules, and their packing are all sensitive to the localization of electrons and the magnitude of the charge imbalance, GGA may not have enough accuracy to calculate the electronic structures of such molecular CO systems. 
Due to this problem of GGA-PBE, it is very difficult to construct a reliable effective model, which is essential for many theoretical studies related to the CO state. 
Some of the parameters in the effective model may be determined from the experimental results, but information from first-principles calculations is highly desirable.

In this work, we compare the GGA-PBE functional with another exchange-correlation functional, i.e., a hybrid functional by Heyd, Scuseria, and Ernzerhof (HSE06),~\cite{HSE03, HSE04, HSE_Martin_JCP, Errata_HSE06} and investigate the stability of the CO states in D-S and D-Se. 
We expect that the hybrid functional method will provide more reliable results than GGA to calculate the electronic and structural properties of this class of materials.~\cite{HSE_TTFCA_PRL2009, HSE_TTF_TCNQ} 
However, since the computational cost is much more expensive than that of GGA, they have not been applied to complex molecular CO systems and their quantitative accuracy is yet to be evaluated.

The rest of the paper is organized as follows. 
We introduce the crystal structure and the calculation method in Secs. II and III, respectively. 
In Sec. IV, we show the results: The electronic structure is discussed based on the experimental structures of D-S and D-Se and the use of GGA-PBE and HSE06 functionals are compared (Sec. IV A). Structural optimization for D-S is performed, and the structural stability of the CO phase is discussed, including the possibility of another CO pattern (Sec. IV B). Sections. V and VI are devoted to discussion and a summary, respectively. 
%%%%    Fig.1    %%%%%%%%%%%%%%%%%%%%%%%%%%%%%%%%%%%%%%%%%%%%%%%%%
%%%%    Fig.1    %%%%%%%%%%%%%%%%%%%%%%%%%%%%%%%%%%%%%%%%%%%%%%%%%
\section{Crystal structure}
The low-temperature phase of D-S and D-Se has a triclinic structure (centrosymmetric space group: $P\bar{1}$) where one of the D atoms is located at an off-center position closer to one of the two O atoms forming hydrogen bonds, as shown in Fig.~\ref{Crystal}(a).
This is in contrast to the high-temperature phase (centrosymmetric space group: $C2/c$), isostructural to the H samples, where all the H/D(Cat-EDT-TTF/ST) units are equivalent and the hydrogen bond-forming H/D is equally shared by two oxygens in different units.  
Figures~\ref{Crystal}(b) and~\ref{Crystal}(c) show the crystal structure, which is characterized by a charge disproportionation between monomers w-D and w/o-D. 
Each monomer forms a dimer with the same type of monomer; namely, there are w-D dimers and w/o-D dimers with a 1:1 ratio. 
These two kinds of dimers, which are in the high-temperature $C$2/$c$ structure equivalent and
connected by glide operations, are alternately stacked with a periodic arrangement. 

\section{Calculation method}
In order to study the electronic and structural properties of the CO state coupled with D ordering, 
we compare the results with two first-principles approaches mentioned above, using GGA-PBE and HSE06 as exchange-correlation functionals. 
Note that we need to treat strong intramolecular and weak intermolecular interactions simultaneously.
Furthermore, the energy difference between competing phases is often very small, and can be as small as of the order of 10 meV/f.u. in our case, as discussed later. 

For the GGA calculations, Kohn-Sham equations are self-consistently solved in a scalar-relativistic fashion using the all-electron full-potential linearized augmented plane
wave~(FLAPW) method implemented in the QMD-FLAPW12 code~\cite{Wimmer1981, L_KA, Weinert, P3HT_Tsumu}.
The band structure calculations with GGA are also performed using the pseudopotential method based on the projector augmented wave (PAW) method with plane wave basis sets implemented in the Vienna $Ab$ $initio$ Simulation Package.~\cite{Vanderbilt1990, Kresse_VASP1996, Kresse_Joubert_PAW1999} 
The results are fairly in agreement with each other.
The LAPW basis functions in the interstitial have a cutoff energy of 30.3 Ry.  
The angular momentum expansion inside the sphere is truncated at $l$ = 8 for all the atoms. 
The cut-off energy for the potential and density is 276 Ry. 
The muffin-tin sphere radii are set as 0.66, 0.36, 0.62, 1.03 and 1.16~\AA~for C, H, O, S, and Se atoms, respectively. 
The $\bf{k}$-point meshes used are 4 $\times$ 4 $\times$ 2. 

As for the HSE06 hybrid functional,~the Kohn-Sham equations are solved using the pseudopotential method based on PAW method with plane wave basis sets implemented in VASP.~\cite{Paier_HSE06}
Previously, the role of the exact exchange on the electronic structure for the charge modulated state was discussed for TiSe$_2$~\cite{Chen2015_CDW_TiSe2, CDW_HSE-TiSe2} and BaBiO$_3$,~\cite{FoyevtsovaHSE_BaBiO3,Boeri_BaBiO3_HSE} and it was shown that the use of the hybrid functional was essential for obtaining a proper description of the electronic and structural properties. 
In the present HSE06 calculations, we first obtain a converged charge density from the self-consistent calculation within GGA, and then the self-consistent hybrid functional calculations are performed using the GGA charge density as the initial state. 
A common $\bf{k}$-point sampling, including for the structural optimization, is set as  3 $\times$ 4 $\times$ 2. 
The cut-off energy for plane waves is 29.4 Ry for the HSE06 calculations. 
The range-separation parameter in HSE06 calculations is 0.2~\AA$^{-1}$, and 25$\%$ of the exact exchange is mixed to the GGA exchange for the short-range interactions. 

We note that the present study does not consider the quantum effects of H or D atoms.~\cite{F_Ishii06_organicFE, Horiuchi2010, Ishibashi2019_FE} 
We do not distinguish H and D atoms in our DFT calculations, and the H/D isotope effect is not discussed. 

For the lattice parameters, we use the experimental ones throughout this paper, since we have difficulty with their theoretical determination as explained in the following. 
First, the importance of dispersion interactions or van dar Waals interactions, 
which are not correctly treated in the GGA-PBE or HSE06 functional, in the CO phase is unclear.~\cite{D3_comment}
We expect the electrostatic interactions in the CO phase are much stronger.~\cite{SeoHott_ChmRev04, Naka_PRB_H3Cat18} 

Second, if the effect of dispersion interactions is important, the choice of 
the method treating the dispersion corrections would significantly affect 
the optimized lattice parameters.
To evaluate the lattice parameters accurately, we need to use a method which 
can treat the dispersion interactions as well as the CO state, simultaneously. 
We may also have a problem in the numerical accuracy to calculate the stress tensors since they 
are subtle quantities related to the calculation conditions. 
Considering these ambiguities in the optimization of lattice parameters and, in contrast, the high accuracy of the experimental ones
especially at ambient pressure, we use the experimental lattice parameters and concentrate on 
clarifying the detailed accuracy of the first-principles evaluation of various quantities related 
to the CO state of the title compounds.
Since the reliability and accuracy of first-principles methods within GGA-PBE or HSE06 for the
CO state in molecular conductors are far from clear, we believe
that our present study is also valuable for constructing a reliable effective model.

\section{Results}
As explained above, it is important to investigate the accuracy of 
the first-principles methods for the title compounds.
In this section, we first investigate the electronic properties of the CO state using the 
experimental structure, including the internal coordinates, in Sec. IV A. 
Then, in Sec. IV B, we work on its structural stability by optimizing the internal coordinates using GGA-PBE and HSE06 functionals.

\subsection{Electronic structure}
\begin{figure}[tb]
\begin{center}
\includegraphics[width=1.0\linewidth]{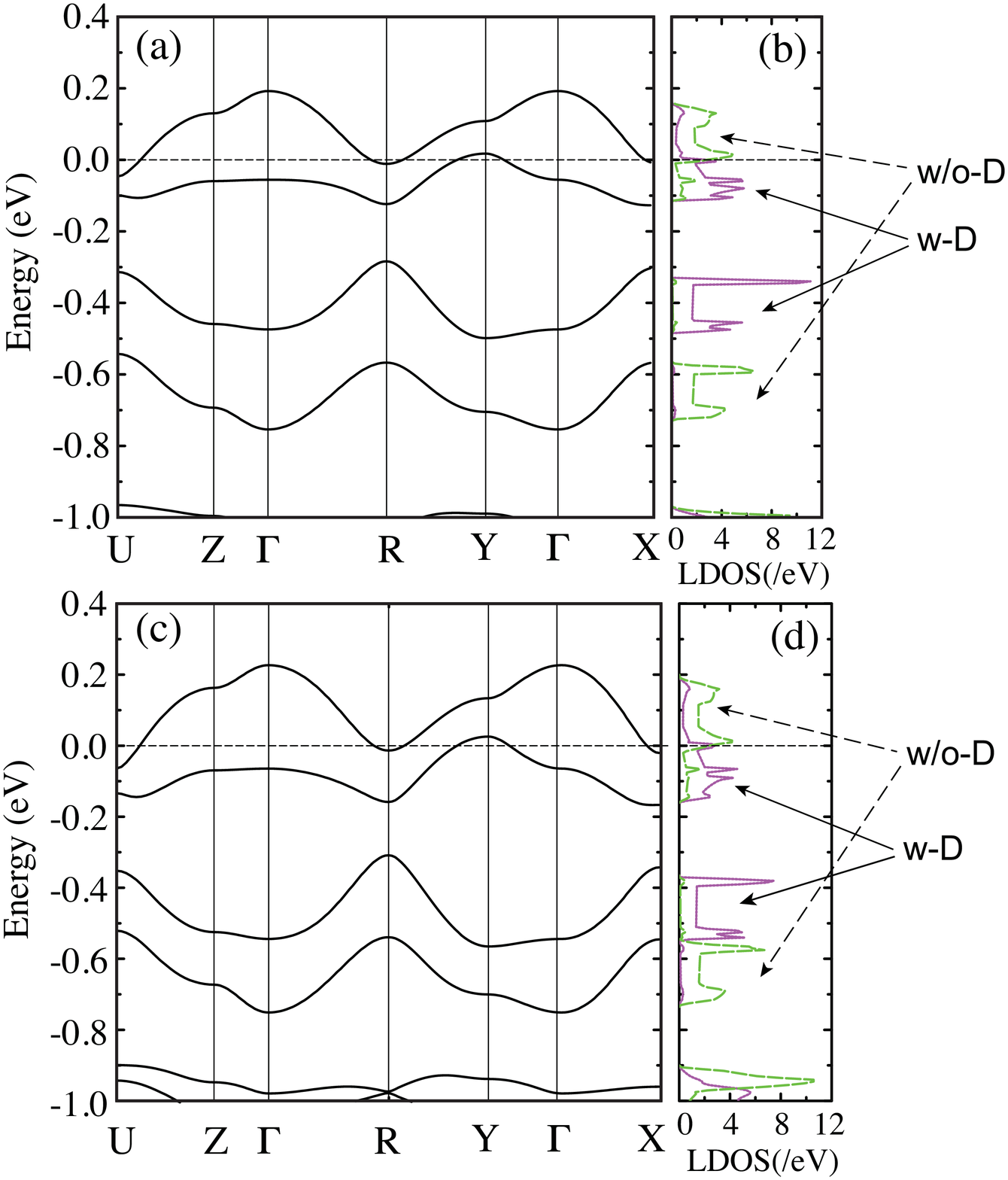}
\end{center}
\caption{(a)~Band structure and (b)~local density of states (LDOS) of the low-temperature phase of $\kappa$-D$_3$(Cat-EDT-TTF)$_2$ calculated within GGA. 
The origin of the vertical axis with a dashed line shows the Fermi level. 
The LDOS contains two parts; the solid and the dashed curves indicate LDOS of D$_2$(Cat-EDT-TTF) (w-D) and D(Cat-EDT-TTF) (w/o-D), respectively.  
(c)~Band structure and (d)~LDOS of the low temperature phase of $\kappa$-D$_3$(Cat-EDT-ST)$_2$.}
\setlength\abovecaptionskip{0pt}
\label{band_H3Cat}
\end{figure}

In this section, the electronic structure based on the experimental crystal structure in the CO state is discussed. 
Figure~\ref{band_H3Cat}(a) shows the calculated band structure of D-S within GGA.   
The top of the valence band is located at the $Y$-point and the bottom of the conduction band is located at the $U$-point, which both slightly cross the Fermi level; the system is (semi-)metallic.
%%%%    Fig.2   \label{band_H3Cat} %%%%%%%%%%%%%%%%%%%%%%%%%%%%%%%%%%%%%%%%%%%%%%%%%
%%%%    Fig.2    %%%%%%%%%%%%%%%%%%%%%%%%%%%%%%%%%%%%%%%%%%%%%%%%%
As plotted in Fig.~\ref{band_H3Cat}(b), the calculated local density of states~(LDOS) clearly shows charge disproportionation between the two types of monomers in the unit cell.  
The solid and dashed curves are the LDOS of the w-D and w/o-D units, respectively.  
The LDOS are obtained as a summation of projected DOS on C $p$, O $p$, S $p$ and S $d$ states in the respective monomer units. 
Compared with the CO state discussed in our previous study for H-S,~\cite{Tsumu_H3Cat}  
 the magnitude of the intermolecular charge disproportionation is much larger, which is consistent with the experimental observations. 
At around -0.1 to 0 eV and --0.5 to --0.3 eV, where the second and third bands locate among the four bands, the LDOS of w-D is much larger than that of w/o-D. 
Conversely, the top and fourth bands at around 0 to +0.2 eV and --0.7 to --0.5 eV, respectively, mostly originate from the LDOS of w/o-D. 

\begin{figure}[tb]
\begin{center}
\includegraphics[width=0.8\linewidth]{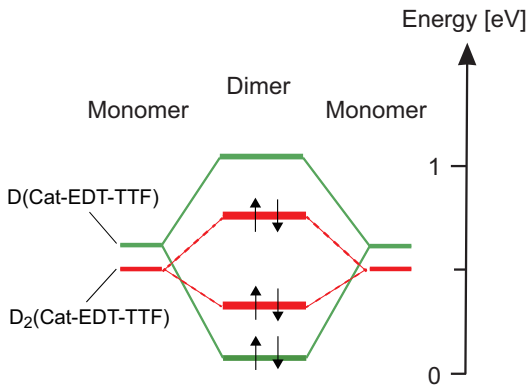}
\end{center}
\caption{(Color online) Schematic energy diagram of the HOMO orbitals for the w-D and w/o-D monomers and their dimers, D$_4$(Cat-EDT-TTF)$_2$ and D$_2$(Cat-EDT-TTF)$_2$, calculated with the GGA-PBE functional. 
The energy diagram with dashed (solid) lines shows that of D$_4$(Cat-EDT-TTF)$_2$~[D$_2$(Cat-EDT-TTF)$_2$].
The HSE06 functional gives almost the same result, but the energy position of the antibonding state of D$_2$(Cat-EDT-TTF)$_2$ is about 0.22 eV higher than that of GGA. Here, the energy levels of the monomers are evaluated by single-$\bf{k}$-point calculations of isolated monomers in a large supercell. 
On the other hand, the MO levels of the dimers are obtained from single-$\bf{k}$-point calculations at the $\Gamma$ point for the crystalline solid of $\kappa$-D$_3$(Cat-EDT-TTF)$_2$.
}
\setlength\abovecaptionskip{0pt}
\label{MO_D3Cat}
\end{figure}

These features can be understood by a schematic energy diagram of molecular orbitals of the two kinds of dimers, as shown in Fig.~\ref{MO_D3Cat}.
Here, the energy levels of monomers are evaluated by the calculations of the isolated monomers in a supercell. 
We first note that the HOMO level of the w/D unit is lower than that of the w/o-D unit. 
This is reasonable since we expect that it is stabilized by the hydroxyl (-OD) group at the end of the molecule. 

When they form dimers, the energy levels are split into antibonding and bonding states, and since the 
distances between the two monomers are very different [see Figs.~\ref{Crystal}(d) and \ref{Crystal}(e)], the splitting is much larger in the w/o-D units with shorter distance (3.42~\AA) than in the w-D units with larger distance~(3.71~\AA). These MO levels of the dimers are obtained from a single-$\bf{k}$-point calculation for the crystalline solid. 
As a result, the antibonding level of the w/o-D dimer is higher than that of the w-D dimer, and the occupancies of the two antibonding levels 
become different, 2 for w-D and 0 for w/o-D dimers, consistent with the realization of the CO state in the solid.
From these analyses, we conclude that there are two main contributions to stabilize the CO state: 
i) the energy difference between w/o-D and w-D monomers, and ii) the stronger dimer formation of w/o-D than w-D. 
Since the former is closely related to the localization of H or D atoms, their positions and dynamics are coupled to the stability of the CO state.

Next, the GGA band structure of D-Se and the LDOS of the w-D and w/o-D units are shown in Figs.~\ref{band_H3Cat}(c) and~\ref{band_H3Cat}(d), respectively.  
Similarly to D-S, the Fermi level crosses the upper two bands resulting in a metallic state, and the LDOS shows charge disproportionation between the w-D and w/o-D units. 
The difference from D-S is that the dispersion of each band is about 15 $\%$ larger than that of D-S while, from the structural point of view, the interplanar distances between monomers are slightly longer~[Figs.~\ref{Crystal}(d) and ~\ref{Crystal}(e)]. 
These originate from the fact that the Se-$p$ state is more delocalized than S-$p$. 
Again, good correspondence between the four bands and the LDOS of the four monomer units is seen. 
Therefore, we expect that the mechanism of CO formation coupled to the D ordering is the same as in D-S. 

%%%%    Fig.3    %%%%%%%%%%%%%%%%%%%%%%%%%%%%%%%%%%%%%%%%%%%%%%%%%
%%%%    Fig.3    %%%%%%%%%%%%%%%%%%%%%%%%%%%%%%%%%%%%%%%%%%%%%%%%%

\begin{figure}[tb]
\begin{center}
\includegraphics[width=1.0\linewidth]{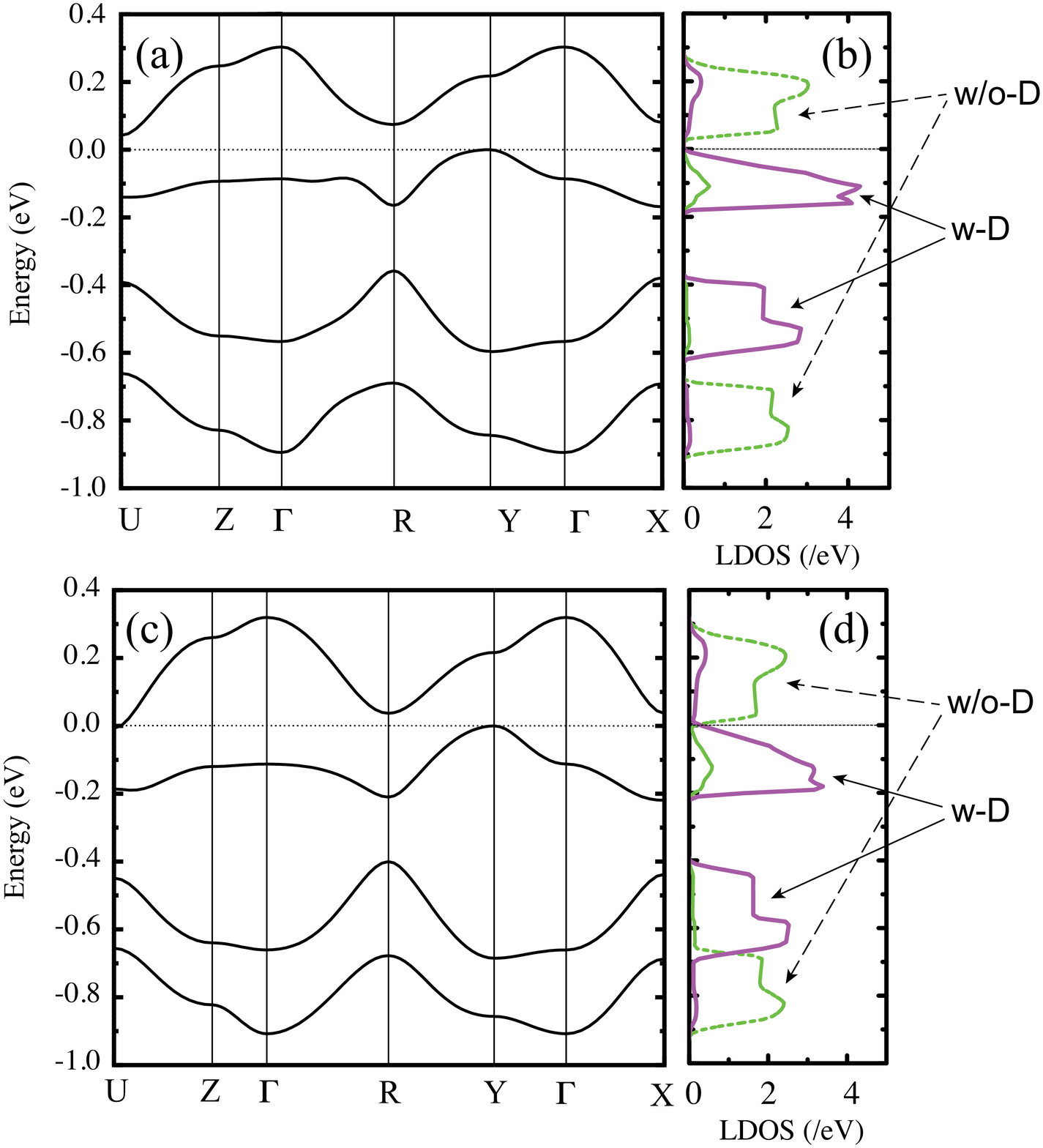}
\end{center}
\caption{(Color online) (a) HSE06 band structure and (b) local density of states~(LDOS) of the low temperature phase of $\kappa$-D$_3$(Cat-EDT-TTF)$_2$. 
(c) HSE06 band structure and (d) LDOS of $\kappa$-D$_3$(Cat-EDT-ST)$_2$. 
Similar to Fig.~\ref{MO_D3Cat}, the plotted LDOS contains two parts; the solid and the dashed curves indicate LDOS of w/D and w/o-D units, respectively.
The dotted lines at 0 eV show the top of the valence bands.}
\setlength\abovecaptionskip{0pt}
\label{bands_TTF_HSE}
\end{figure}

Now, the band structures and the LDOS of D-S and D-Se using the HSE06 hybrid functional are plotted in Fig.~\ref{bands_TTF_HSE}. 
Compared to the GGA results, the four bands are farther apart from each other, suggesting the more localized behavior of the wave functions, as expected. Since the same correspondence between the bands and the molecular orbital levels discussed above holds, the enhanced separation of the bands leads to more pronounced charge disproportionation. This is seen in the LDOS in Figs.~\ref{bands_TTF_HSE}(b) and~\ref{bands_TTF_HSE}(d), especially in the upper two bands where the LDOS of the antibonding state of w-D (w/o-D) units becomes more occupied (unoccupied) compared to the GGA case in Fig.~\ref{band_H3Cat}. 

In D-S, more importantly, a finite band gap is opened. An indirect band gap of 0.04 eV is obtained between the top of the valence bands at the $Y$-point and the bottom of conduction bands at the $U$-point.
The same tendency is seen in the band structure of D-Se, but it is not enough to make the system an insulator. 
Although there are some quantitative differences, GGA and HSE06 provide similar band structures for both compounds. However, we will show that the structural stability of the CO state is completely different between these two functionals, in the next section. 

%%%%    Fig.4    %%%%%%%%%%%%%%%%%%%%%%%%%%%%%%%%%%%%%%%%%%%%%%%%%
%%%%    Fig.4    %%%%%%%%%%%%%%%%%%%%%%%%%%%%%%%%%%%%%%%%%%%%%%%%%

\subsection{Structural optimization and possible noncentrosymmetric phase}
In this section, we investigate the accuracy of structure determination of the internal coordinates using the experimental lattice parameters, as in our previous work.~\cite{Tsumu_H3Cat} 
First, we discuss the results obtained by setting the initial state as the experimental structures of the D localized phase in D-S and D-Se. 
We have also found another stable structure with noncentrosymmetric space group $P$1, which will be discussed later. 

\begin{figure}[tb]
\begin{center}
\includegraphics[width=1.0\linewidth]{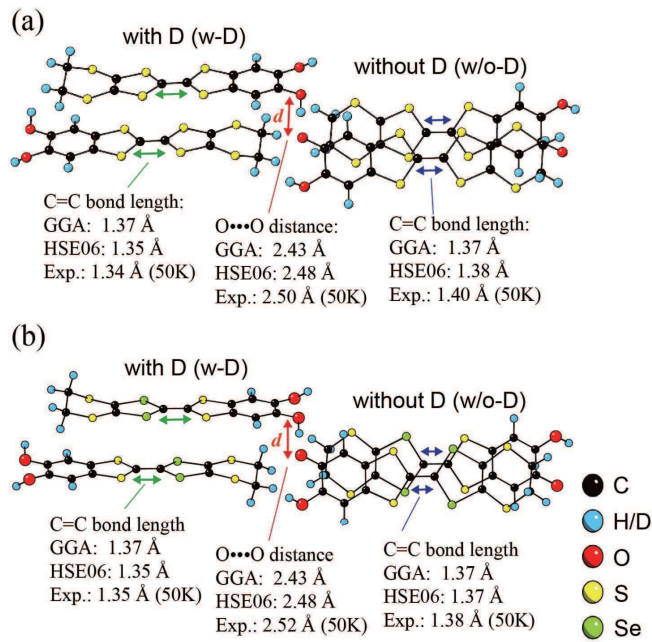}
\end{center}
\caption{(Color online) Optimized bond length of the C=C bond at the center of the TTF part and the O$\cdots$O distance at the hydrogen bonding in (a) $\kappa$-D$_3$(Cat-EDT-TTF)$_2$ and (b) $\kappa$-D$_3$(Cat-EDT-ST)$_2$, calculated with the GGA-PBE and HSE06 functionals, together with their experimental values.}
\setlength\abovecaptionskip{0pt}
\label{Structure_HSE}
\end{figure}

The experimental and theoretically optimized bond parameters are summarized in Table~\ref{OHgeo}, and the O$\cdots$O and C=C distances are shown in Fig.~\ref{Structure_HSE}. 
The first thing we notice is that, after the geometrical relaxations within GGA-PBE, the difference in the C=C bond length at the center of the TTF part of w-D and w/o-D is less than 10$^{-3}$\AA, in both D-S and D-Se. 
This is also the case for the rest of the bond lengths in the two monomer units, suggesting that the structure is basically relaxed to that of higher symmetry $C$2/$c$ of the high-temperature phase, within the numerical accuracy. 
Since this C=C bond length is known to be sensitive to the charge
disproportionation, the results indicate that CO state is unstable after the structural optimization.

We can overcome this problem using the HSE06 hybrid functional.  
Now, as seen in Table~\ref{OHgeo} and Fig.~\ref{Structure_HSE}, the optimized structures are fairly in agreement with the experiments. 
Importantly, the central C=C bond lengths in the TTF part are well reproduced, 1.35 and 1.38 \AA~for the w-D and w/o-D units (for both D-S and D-Se) respectively, within $\pm$ 0.02 \AA~compared with the experimental values. 
Therefore, we conclude that the HSE06 functional is highly accurate for reproducing the structures of the molecular CO phase quantitatively. 

%%%%    Fig.5    %%%%%%%%%%%%%%%%%%%%%%%%%%%%%%%%%%%%%%%%%%%%%%%%%
%%%%    Fig.5    %%%%%%%%%%%%%%%%%%%%%%%%%%%%%%%%%%%%%%%%%%%%%%%%%

\begin{table*}[tb]
\caption{Experimental and theoretically optimized structural parameters of central C=C bonds in the TTF part and hydrogen bonds in $\kappa$-D$_3$(Cat-EDT-TTF)$_2$ (D-S) and $\kappa$-D$_3$(Cat-EDT-ST)$_2$ (D-Se). The $P$1 structure for D-S is the noncentrosymmetric structure discussed in the text.}
\label{OHgeo}
\begin{center}
\begin{tabular}{llccccccc}
\hline\hline
 & Method & C=C~[\AA] & C=C~[\AA] & O$\cdots$O~[\AA] & O--D~[\AA] & O--D~[\AA] & $\angle$O--D--O~[deg] \\ 
 &       & w-D & w/o-D &  & w-D & w/o-D &  \\

\hline
D-S \\
Centrosymmetric~($P\bar{1}$) & Exp. & 1.34 & 1.40 & 2.50 & 1.02 & 1.51 & 164.7 \\
 & GGA-PBE & 1.37 & 1.37 & 2.43 & 1.14 & 1.29 & 173.9 \\
 & HSE06 & 1.35 & 1.38 & 2.48 & 1.06 & 1.41 & 173.9 \\ 
 & & & & & & \\
Noncentrosymmetric~($P1$) & HSE06 & 1.37 & 1.38 & 2.43 & 1.10 & 1.33 & 174.7 \\ 
 &            & 1.35 & 1.36 & 2.49 & 1.05 & 1.43 & 173.3 \\ 
\hline
D-Se \\
Centrosymmetric~($P\bar{1}$) & Exp.\cite{A_Ueda_DCat_Eur} & 1.35 & 1.38 & 2.52 & 0.80 & 1.73 & 167.9\\
 & GGA-PBE & 1.37 & 1.37 & 2.43 & 1.09 & 1.40 & 172.9 \\ 
  & HSE06 & 1.35 & 1.38 & 2.48 & 1.06 & 1.43 & 173.8 \\ 
\hline\hline
\end{tabular}
\end{center}
\end{table*} 

Encouraged by the reliability of the results above, we now investigate another CO state with D ordering, but with a different pattern from the experimental observations. We consider the CO pattern known as the ``ferroelectric'' phase discussed in $\kappa$-(BEDT-TTF)$_2$$X$~\cite{BEDT-TTF} and related systems.~\cite{NakaIshihara, Majed2010, Gomi2010, Naka_JPSJ_2013, Gomi_FE2013, Gati_ET_FE_2018, Hassan1101} It gives rise to a noncentrosymmetric $P$1 structure, and its possible realization coupled to the displacement of H/D in H/D-S has been discussed in Ref.~\cite{Naka_PRB_H3Cat18} based on an effective model. 

\begin{figure*}[tb]
\begin{center}
\includegraphics[width=0.85\linewidth]{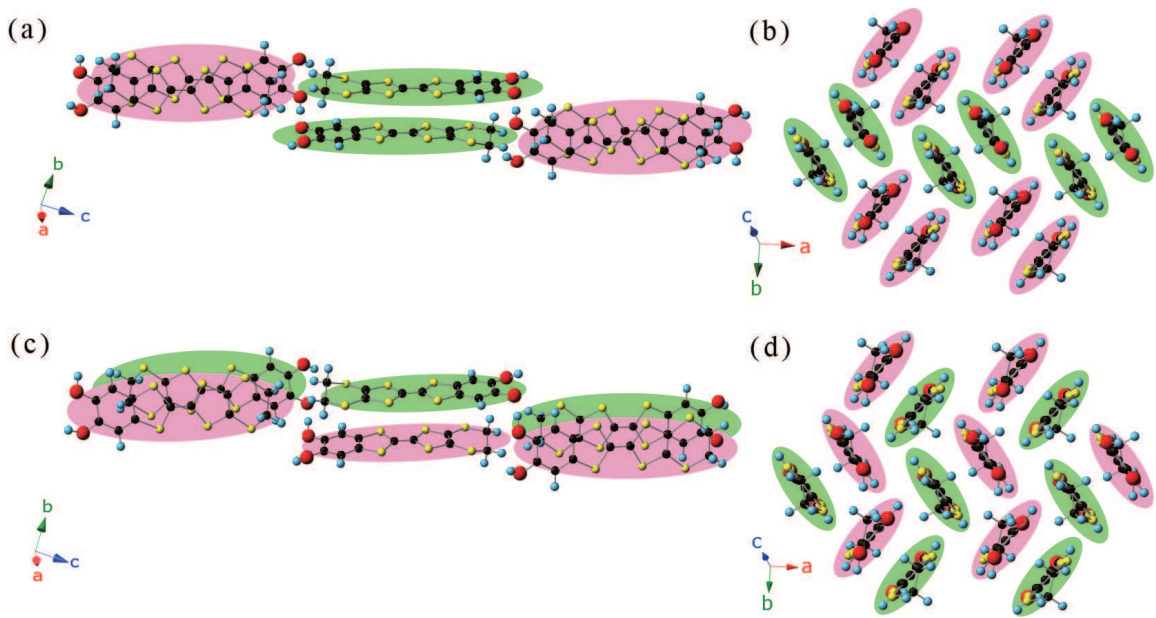}
\end{center}
\caption{(Color online) Two different ordering patterns of the shared D atom in D-S: (a)~Centrosymmetric ($P\bar{1}$) structure and (b) a view along the $c$ axis showing the $ab$ plane [the same as in Figs.~1(b) and (c) but with expanded views]. (c) Noncentrosymmetric ($P1$) structure with a different D ordering pattern, and (d) a view along the $c$ axis showing the $ab$ plane. 
The structures optimized with the HSE06 hybrid functional are shown.}
\setlength\abovecaptionskip{0pt}
\label{P1barP1_HSE_opt}
\end{figure*}

Here, to model the noncentrosymmetric phase, in the initial state we change the position of D atoms,
from the experimental structure of the low-temperature phase of D samples,
to that where the direction of the displacement of D from the middle point between the shared oxygens is opposite. 
We indeed find a stable solution by structural optimization, as listed in Table~\ref{OHgeo}. 
One can see that there are four kinds of C=C bond lengths as well as O-D distances, which we can classify into two each: the w-D (w/o-D) units with shorter (longer) C=C bonds, indicating larger (smaller) electron occupation. 
The pattern is shown in Fig.~\ref{P1barP1_HSE_opt}, together with the experimentally observed pattern [the same as in Figs.~\ref{Crystal}(b) and (c) but with expanded views]. 
In the latter [Figs.~\ref{P1barP1_HSE_opt}(a) and \ref{P1barP1_HSE_opt}(b)], the two kinds of dimers, i.e., the w-D dimers and w/o-D dimers are arranged in a checkerboard pattern in the two-dimensional planes. 
On the other hand, as shown in~Figs.~\ref{P1barP1_HSE_opt}(c)~and~(d), in the noncentrosymmetric phase, 
each dimer has one w-D unit and one w/o-D unit, leading to charge disproportionation and electric polarization within the dimers. 
The pattern globally breaks the inversion symmetry.  

Energetically,~ the centrosymmetric and noncentrosymmetric structures are very close, while the centrosymmetric phase~(-406.190~eV/f.u.) is slightly more favorable than that of the noncentrosymmetric structure~(-406.182~eV/f.u.) within an energy difference of 8~meV per formula unit.~
It is an interesting issue to evaluate the value of electrical polarization and identify its nature in the noncentrosymmetric phase~\cite{Horiuchi2010, King-Smith93, Resta94, Ishibashi_FE2010}, under the variation from a ``paraelectric" phase involving both D displacements and molecular deformation, which we leave as a future problem. 

%%%%    Fig.6    %%%%%%%%%%%%%%%%%%%%%%%%%%%%%%%%%%%%%%%%%%%%%%%%%
%%%%    Fig.6    %%%%%%%%%%%%%%%%%%%%%%%%%%%%%%%%%%%%%%%%%%%%%%%%%

\section{Discussion}
Using the HSE06 functional, our theoretical results by structural optimization show overall good agreement with the experimental structural parameters of the low-temperature phase in D-S and D-Se.
This indicates that the more localized nature of the resulting wave functions is reflected in the electron-lattice coupling giving rise to a quantitative description of the materials. 
In TTF-based systems, it is known that the amount of charge on the molecule can be probed by measuring the frequency of the central C=C bond vibration by Raman vibrational spectroscopy.~\cite{Yue_Kaoru_Y_alpha_2010, Hassan1101}
In fact, the low-temperature Raman spectrum of D-S shows new peaks at 1516 and 1405~cm$^{-1}$ that are not observed in the high-temperature phase.~\cite{A_Ueda_DCat} 
These are assigned to the shorter and longer central C=C bonds of the w-D and w/o-D units, respectively.
With a reliable description of the CO phase as demonstrated here, 
we expect that the phonon frequencies calculated on the basis of our results can now be directly compared with experiments.~\cite{Tsumuraya_2009} 

The band gap is opened in our HSE06 calculations for D-S, and its value is in fairly good agreement with the experimental value estimated from the resistivity measurement.~\cite{A_Ueda_DCat, A_Ueda_DCat_Eur} 
On the other hand, for D-Se, our calculations exhibit a semi-metallic band structure while experiments show an insulating behavior.
The full clarification of the accuracy of HSE06 for the evaluation of band gaps in strongly correlated molecular systems is left for future studies.

Finally let us discuss our results in light of previous theoretical studies for these compounds. 
Multi-component DFT based quantum chemistry calculations were performed by Yamamoto {\it et al.}~\cite{KaichiY_H3Cat, KYamamoto_H3Cat_ST} with a hybrid functional, the so-called Minnesota functional.
The nuclear quantum effect of H/D on the potential energy surface of the H/D atoms was studied: It changes the potential surface from double well to single well.
We consider that the two DFT-based calculation methods are complementary to each other: 
In our calculations, the quantum effect of H/D atom cannot be discussed, while the structural stability of the low-temperature phase in the actual crystalline form can be investigated in contrast to their calculations where only isolated molecular systems are treated. 

On the other hand, Naka and Ishihara~\cite{Naka_PRB_H3Cat18} studied an effective model where the correlated $\pi$-electron system is coupled with pseudospins describing the proton~(H$^+$) degree of freedom, assuming its double well potential. Their results show that two types of $\pi$-electron-proton coupled order compete with the dimer-Mott insulating state, both resulting from the cooperative effects of the proton ordering and CO. The two ordered states corresponds to the centrosymmetric and noncentrosymmetric phases discussed above.  
They considered the proton-electron coupling as an attractive potential, which is consistent with our results that the w-D monomers show lower HOMO energy than the w/o-D monomers (Fig.~\ref{MO_D3Cat}). However, in their model, the difference of inter-monomer distances in each dimer is not taken into account. 
Our results suggest that the electronic structure of the low-temperature phase is stabilized by both effects, namely, the variation of the intermolecular distances in two different dimers and the difference in MO energy levels depending on the position of the D atom. 
By the quantitative evaluation using first-principles calculations, these two effects indeed have the same order, as seen from our energy diagram in Fig.~\ref{MO_D3Cat}. 

\section{Summary}
We studied the electronic and structural properties of the deuterium and $\pi$-electron coupled CO state of $\kappa$-D$_3$(Cat-EDT-TTF)$_2$ (D-S) and $\kappa$-D$_3$(Cat-EDT-ST)$_2$ (D-Se) using GGA and a range-separated hybrid functional HSE06. 
Using the experimental crystal structure, both GGA and HSE06 calculations show charge ordering but with a larger degree of charge disproportion in the latter, closer to the experimental situation. 
Using HSE06, an insulating band structure of D-S is obtained.  
By performing structural optimization with HSE06, the central C=C bond lengths that are sensitive to the degree of charge disproportionation are well reproduced, whereas GGA fails to stabilize the CO state. 
We also proposed possible patterns of D ordering and showed a stable noncentrosymmetric CO phase which has energy close to the experimentally realized phase.  

\section{ACKNOWLEDGEMENTS}
The authors thank A.~Ueda, H.~Mori, M. Naka, H. Watanabe, T.~Sasaki, and R. Kato~for stimulating discussions. 
This research was funded by JSPS Grant-in-Aid for Scientific Research Nos.~16K17756, 26400377, 19K04988, 19K03723, 19K21860 and JST CREST Grant No.~JPMJCR18I2. 
One the authors (T.T.) is partially supported by MEXT Japan, Leading Initiative for Excellent Young Researchers (LEADER). 
Computational work was performed under the Inter-university Cooperative Research Program and the Supercomputing Consortium for Computational Materials Science of the Institute for Materials Research (IMR), Tohoku University (Proposal No. 18K0090 and 19K0043). The computations were mainly carried
out using the computer facilities of ITO at the Research Institute for Information Technology, Kyushu University, MASAMUNE at IMR, Tohoku University, NIMS Material Simulator, and HOKUSAI-GreatWave at RIKEN.

%\newpage
\bibliography{./Tsumu}
\end{document}